# ALFVEN WAVE GENERATION BY MEANS OF HIGH ORBITAL INJECTION OF BARIUM CLOUD IN MAGNETOSPHERE

V.N. Oraevsky, Yu.Ya. Ruzhin, V.I. Badin, M.G. Deminov

*Institute of Terrestrial Magnetism, Ionosphere and Radiowave Propagation, Troitsk, Moscow Region, 142190, Russia*

## ABSTRACT

An analysis of the Alfven wave generation associated with the barium vapor release at altitudes ~ 5.2 $R_E$ in the magnetosphere is presented. Such injections were executed in G-8 and G-10 experiments of the Combined Radiation and Radiation Effects Satellite (CRRES) mission. It is shown that the generation of Alfven waves is possible during the total time of the expansion of plasma cloud. The maximum intensity of these waves corresponds to the time of complete retardation of the diamagnetic cavity created by the expansion of the plasma cloud. The Alfven wave exhibits a form of an impulse with an effective frequency ~ 0.03-0.05 Hz. Due to the background conditions and wave frequency, the wave mainly oscillates along the geomagnetic field between the mirror reflection points situated at ~ 0.7 $R_E$ i.e. the wave is trapped by the magnetospheric Alfven resonator. The reflection coefficient is about 80-85%. The wave amplitude is sufficient to the generation of plasma instabilities and longitudinal electric field, and to an increase in the longitudinal energy of electrons accelerated by this field up to ~ 1 keV. These processes are the most probable for altitudes ~ 1 $R_E$. The auroral kilometric radiation (AKR) at frequencies ~ 100 kHz is associated with these accelerated electrons. It is shown that the acceleration of electrons and AKR can be observed almost continuously during the first minute and then from time to time with pauses about 35-40 s till 6-8 min after the release. Apparently, these findings do not contradict the experimental data. The betatron acceleration of electrons at the recovery of the geomagnetic field is also discussed. This mechanism could be responsible for the acceleration of electrons resulting in the aurorae and ultra short radio wave storm at frequencies 50-300 MHz observed at the 8-10th min after the release.

INTRODUCTION

The atomic barium releases at altitudes ~ 5.2 $R_E$ in the magnetosphere were executed in G-8 (February 17, 1991) and G-10 (January 20,1991) experiments of the Combined Radiation and Radiation Effects Satellite (CRRES) mission (Bernhardt,1992). A number of effects resulting from the expansion and ionization of the barium cloud were observed: the creation of a diamagnetic cavity with the background magnetic field almost totally excluded; formation of irregularities on the surface of the diamagnetic cavity and in the plasma cloud; increase in the auroral activity (for G-10: very faint rays upon 4 min after the release and a more distinct increase in the auroral activity upon 9-10 min after the release); auroral kilometric radiation (AKR) with frequencies about 100 kHz that apparently was observed for a long time (Bernhardt, 1992). During the G-8 experiment, an ultra short radio wave storm at frequencies 45-90, 235, and 280 MHz started upon approximately 10 min after the release (Oraevsky et al., 1995). Huba et al. (1992) carefully studied the processes of formation of the diamagnetic cavity and large-scaled plasma irregularities and the results of their theoretical analysis well agree with the experimental data. Oraevsky et al. (1995) proposed theoretical estimates for the possible mechanism of the ultra short radio wave storm.

The main goal of this work is to present a qualitative study of the possibility of the Alfven wave generation at the expansion of the plasma cloud in the magnetosphere and the estimates of some plausible effects associated with this wave. In addition, the possibility of plasma heating by the collapse of diamagnetic cavity is qualitatively discussed since the electrons with high energies are necessary for an explanation of the auroral activity and ultra short radio wave storm. Our study and estimates are qualitative and preliminary formulating rather a hypothesis than a strict result.

GENERATION OF THE ALFVEN WAVE

We use the parameters relevant to the experiment G-10 when at the time $t = 0$ (03:30 UT January 20, 1991), the $N_0 = 1.86 \; 10^{25}$ barium atoms ($M_0 = mN_0 = 4240$ g) were injected into the magnetospheric plasma at the height $h = 5.2 \; R_E$ (8.9N, 75.6W, 33179 km) [Bernhardt, 1992; Huba et al.,1992]. The background conditions for this height were as follows: the geomagnetic field $B_0 = 1.35 \; 10^{-3}$ G, number densities of thermal electrons 2 cm$^{-3}$ and hydrogen atoms ~ 50 cm$^{-3}$, Alfven speed $V_A = B_0/(4\pi\rho)^{1/2} = 2 \; 10^8$ cm/s, ion gyrofrequency $\Omega(H^+) = 13$ rad/s, gyroperiod $\tau(H^+) = 2\pi/\Omega(H^+) = 0.5$ s, ion thermal velocity $v_T(H^+) = 9 \; 10^5$ cm/s, Larmor radius $r(H^+) = 6 \; 10^4$ cm, collision frequency $\nu(H^+, H) = 4 \; 10^{-8}$ s$^{-1}$, electron thermal velocity $v_{Te} = 4 \; 10^7$ cm/s. The densities of background ions and atoms are extremely low and we can assume that the barium cloud expands into vacuum until its total ionization. We neglect the orbital velocity of the satellite and assume that the barium cloud is spherical with the radius $R_0 = V_0 t$ where $V_0 = 1.2 \; 10^5$ cm/s is the radial velocity of the atomic barium. We express the time variations of the $n(Ba)$ and $n(Ba^+)$ in a very approximate form

$$n(\text{Ba}) = N_0 \exp(-t/\tau_{ion}) / [(4/3)\pi R_0^3], \quad (1)$$

$$dn(\text{Ba}^+)/dt = (1/\tau_{ion}) n(\text{Ba}), \quad (2)$$

where $\tau_{ion} = 28$ s is the ionization time scale for barium. The $B_0$, $V_0$, and $\tau_{ion}$ presented here are the same as those used by Huba et al. (1992). The initial velocity of the Ba$^+$ is $V_0$. The barium parameters for the background magnetic field are the following: the gyrofrequency $\Omega(\text{Ba}^+) = 9.5 \ 10^{-2}$ rad/s, gyroperiod $\tau(\text{Ba}^+) = 66.4$ s, Larmor radius $r(\text{Ba}^+) = V_0/\Omega(\text{Ba}^+) = 1.3 \ 10^6$ cm. For these Ba$^+$ parameters, the density of the kinetic energy of cloud ions exceeds the density of the geomagnetic field energy upon a few seconds after the release and a diamagnetic cavity with the background magnetic field almost totally excluded is created. Huba et al. (1992) have studied in detail the dynamics of the diamagnetic cavity, evolution of the ion velocity, and formation of the large-scaled plasma irregularities. Here we neglect these irregularities and the time variation of the mean radial velocity of ions takes the form

$$V = V_0 (1 - B_0^2/[4\pi m(\text{Ba}^+) n(\text{Ba}^+) V_0^2])^{1/2} = V_0 (1 - C_1 t^3/[1 - \exp(-t/\tau_{ion})])^{1/2}, \quad (3)$$

where $C_1 = V_0 B_0^2/3M_0$. This equation follows from the energy conservation law (Huba et al., 1992)

$$m(\text{Ba}^+) n(\text{Ba}^+) V^2 + B_0^2/4\pi = m(\text{Ba}^+) n(\text{Ba}^+) V_0^2.$$

The Eq. 3 shows that the velocity $V$ decreases with time. This indicates that the kinetic energy of ions is spent for the creation of diamagnetic cavity. The radius of this cavity obeys the equation

$$dR/dt = V. \quad (4)$$

We can see from the Eqs. (3) and (4) that, at a certain time $t = t_B$, the velocity $V$ becomes zero and the cavity radius reaches its maximum $R = R_B$. If $t_B \gg \tau_{ion}$, $t_B = 38.7$ s and $R_B = 46.5$ km. These values are overestimated since $t_B$ is in fact comparable to $\tau_{ion}$. Note that almost only the newly created ions with the velocity $V_0$ provide an expansion of the diamagnetic cavity. However, the cavity already generated decays slowly. Therefore, the Eq. (3) for a given time t contains $n(\text{Ba}^+)$ obtained by the integration of the Eq. (2) with respect to time from 0 to $t$. Due to the slow cavity decay, we can approximately assume that $V = 0$ and $R = R_B$ during a certain time interval following the complete retardation. Apparently, this time interval is not smaller than 10-20 s.

Let us consider a spherical layer of the thickness $\Delta R = R_0 - R$. Note that the magnetic field in this layer exceeds the background field since the magnetic field lines "flow about" the diamagnetic cavity and "escape" into the magnetosphere (see Figure 1a). Assume that the magnetic field in the spherical layer equals $2B_0$. The boundary of the diamagnetic cavity and the "flowing about" magnetic lines move in radial direction with the velocity $V$. For this reason, the radial velocity of magnetized electrons equals V as well. The initial radial velocity of ions is $V_0$. Thus, a radial current $J_R$ proportional to $V_0 - V$ arises. For a fixed time $t$, this current is produced by ions and electrons originated over the time interval from $(t - \tau_R)$ to $t$ where $\tau_R = \tau(Ba^+)/8 = 4$ s since this interval approximately corresponds to the strictly radial Ba$^+$ motion immediately after ionization. The motions of the rest of ions are non-correlated with each other and yield no contribution to the radial current. For the initial period of expansion, the plasma kinetic energy is much greater than the energy of the magnetic field forced out by the cloud: $V \approx V_0$, $R \approx R_0$, and almost all currents are closed inside the cloud. To take this into consideration, we assume

$$\tau_R = (1 - V/V_0)\, \tau(Ba^+)/8 \,. \tag{5}$$

The $\tau_R$ reaches its maximum at $V = 0$. Taking into account (2) and (5), we can express the radial current in the form

$$J_R(t) = e\,(1/\tau_{ion}) \int_{t-\tau_R}^{t} (V_0 - V)\, n(Ba)\, dt \,, \tag{6}$$

where the Eq. (1) and (3) determine the $n(Ba)$ and $V$ as explicit functions of time. We can find using these functions that for a relatively short time interval after the release, the integrand is a slow function of time: $(V_0 - V) \sim 0.5 V_0 C_1 t^3$, $n(Ba) \sim 1/t^3$. For this interval, $J_R$ depends on time almost linearly. Near the moment of complete retardation $t = t_B$, the current $J_R$ sharply increases due to an increase in $\tau_R$ and $(V_0-V)$ mainly. For $t > t_B$, this current decreases being proportional to $1/t^3$. We can see these variations of $V$, $R$, and $J_R$ in Figure 1b. This plot shows that $t_B = 34.5$ s, $R_B = 33.6$ km. The time interval $34.5 \le t \le 39$ s corresponds to the maximum of the radial current $J_R = J_{Rm} = 4 \div 4.5 \times 10^{-7}$ A/m$^2$. The duration of this current impulse at the level of $0.5 \div 0.75\, J_{Rm}$ is $\Delta t = 10\text{-}15$ s. This permits to estimate the effective period and frequency of the Alfven wave impulse: $T = 20\text{-}30$ s, $f = 1/T = 0.03\text{-}0.05$ Hz, $\omega = 0.2\text{-}0.3$ rad/s. Below we consider the $J_R$ impulse in detail. Note that the quantities $V$, $R$, and $J_R$ are estimated under an assumption on the homogeneous density distribution of barium atoms inside the sphere of the radius $R_0$. For a more realistic Gauss distribution, the decrease in the density maximum is slower: $t_B \approx 50$ s, $R_B \approx 37$ km (Huba et al., 1992); and we can expect that the $J_R$ current will reach its maximum at a later time after the release.

The currents $J_R$ arising in the spherical plasma layer (envelope) will generate the azimuthal currents $J_\Phi$ and the currents $J_\Theta$. We assume an azimuthal symmetry of the plasma envelope and, consequently, the currents $J_\Phi$ do not contribute to the current continuity div$\mathbf{J} = 0$. In addition, the contribution of $J_\Phi$ is taken into account by the implication of magnetic field changes. For this reason, we consider the currents $J_R$ and $J_\Theta$ only. In this case, the currents $J_\Theta$ flow along the magnetic field (see Figure 1a where the lower hemisphere of the plasma envelope is shown for $0.5\pi < \Theta < 1.5\pi$). The currents $J_\Theta$ do not close inside the plasma layer and transit (near $\Theta \approx \pi$) into the longitudinal currents $J_z$ which flow out from the envelope. Outside the plasma layer, we assume a cylindrical geometry of the magnetic field with coordinates $r, \varphi$, and $z$ where the $z$-axis is parallel to the magnetic field and positive outward the plasma cloud. The currents $J_z$ are the longitudinal currents of the Alfven wave. The $J_R$ is an external current source for the Alfven wave. This $J_R$ current is closed by the radial $J_r$ current of the Alfven wave through the longitudinal currents (see Figure 1a). Near the region of generation, the Alfven wave occupies a cylinder of the radius $r = r_C \approx R_0 - R$ and the amplitude of the longitudinal current of this wave is $J_m \approx J_R$. Here we take into account the current continuity and that the magnetic field outside the plasma layer is equal to the background field $B_0$. We can estimate the amplitudes of the magnetic $B_\varphi$ and electric $E_r$ fields of the Alfven wave near the generation region

$$J_m \approx J_R ; \quad B_m \approx (2\pi/c) J_m r_C ; \quad E_m = (V_A/c) B_m , \qquad (7)$$

where the $J_R$ is determined by the Eq. (6), $V_A$ is the velocity of propagation of this wave. For fixed $z$ and $t$, the longitudinal current $J_z$ integrated from 0 to $r_C$ must yield zero since the Alfven wave does not carry free charges. The simple functions obeying this condition and the Alfven wave equations are as follows:

$$J_z = -J_m (1 - 3 x^2); \quad B_\varphi = -2.6 B_m x (1 - x^2) ; \quad E_r = -2.6 E_m x (1 - x^2) , \qquad (8)$$

where $x = r/r_C$ and $J_m$, $B_m$, and $E_m$ amplitudes are determined by the Eqs. (7). Figure 1c shows the behavior of the functions (8). We can see that it agrees with the qualitative scheme of currents in Figure 1a: the longitudinal currents flow into the cloud ($J_z < 0$) in the central region of the cylinder and flow out ($J_z > 0$) in its peripheral region. The $J_m$ amplitude corresponds to the current flowing into the cloud. For the current flowing out, the amplitude equals $2J_m$. The formulae (8) show that the scale length for changes in $B_\varphi$ and $E_r$ is approximately equal to $0.5\ r_C$. We take this fact into account in the Eqs. (7) for $B_m$ and $E_m$. Since $J_m \approx J_R$, the $J_m$ depends on time (near the generation region) according to Figure 1b where the maximum $J_m$ corresponds to the interval $34.5 < t < 39$ s. The maximum $B_m$ corresponds to the end of this interval ($t \approx 38 \div 39$ s) when the $B_m \approx 3 \div 4\ 10^{-5}$ G and $\delta B_m = B_m/B_0 \approx 2 \div 3\ 10^{-2}$. For the same time, $E_m \approx 6 \div 7$ mV/m in the background plasma.

Thereby, the expanding plasma envelope is a source of the Alfven waves. The maximum intensity of these waves corresponds to the time that immediately follows the complete retardation of the diamagnetic cavity. This maximum exhibits the form of an impulse with the following parameters in the background magnetosphere near the region of generation: the effective frequency $\omega \approx 0.2 \div 0.3$ rad/s, amplitude of the longitudinal current $J_z \approx 4 \div 4.5\ 10^{-7}$ A/m$^2$, magnetic field amplitude $B_\varphi \approx 3 \div 4\ 10^{-5}$ G ($B_\varphi/B_0 \approx 2 \div 3\ 10^{-2}$), and electric field amplitude $E_r \approx 6 \div 7$ mV/m.

PROPAGATION OF THE ALFVEN WAVE

We still use the parameters relevant to the experiment G-10 when the atomic barium was released at the altitude h = 5.2 $R_E$ in the northern hemisphere. This point is situated on the L-shell 7.2 i.e. the apex of the magnetic field line corresponds to the height $h = 6.2\ R_E$. Let the $h_1$ be the height $5.2\ R_E$ in the southern hemisphere, the $h_N = h_S$ be the height $0.7 R_E$ in the northern and southern hemispheres respectively, with the densities of hydrogen and oxygen ions being equal to each other at $0.7 R_E$: $m(H^+)n(H^+) = m(O^+)n(O^+)$. The heights $h_N$ and $h_S$ are selected since the Alfven speed reaches its maximum there: $n_e \approx 15$ cm$^{-3}$, $B \approx 0.1$G, $V_A \approx 6.5\ 10^9$ cm/s. The corresponding distances along this magnetic field line are as follows: $h_1 - h_0$

= $5.72R_E$, $h_N - h_0 = h_S - h_1 = 5.34R_E$. The total length of this field line is approximately equal to 17.9 $R_E$. Below $h_N$ and $h_S$, the speed $V_A$ rapidly decreases and reaches its minimum at the height of the F2 peak of the ionosphere (hmF2 ≈ 300 km): $V_A \approx$ 3-4 $10^7$ cm/s. The Alfven wave propagating upward from altitudes $h <$ hmF2 can be reflected by the upper wall of the ionospheric Alfven resonator (IAR) i.e. the height interval hmF2 $< h < h_N$ and then reflected by the ionospheric dynamo-region ($h$ = 120 km). Thus the wave can be trapped by the IAR (Polyakov and Rapoport, 1981). Note that the Alfven wave can propagate freely when the $V_A$ decreases with propagation and can be reflected from the region where the $V_A$ increases with propagation.

For our case, there is a similar magnetospheric Alfven resonator (MAR) situated on the field line between $h_S$ and $h_N$. The Alfven speed takes its minimum ($V_A \approx$ 9 $10^7$ cm/s) at the apex of the field line taking its maximum ($V_A = V_{Am} \approx$ 6.5 $10^9$ cm/s) at $h_N$ and $h_S$. The Alfven wave is generated inside the MAR and can be trapped by this resonator. The reflection coefficient for the boundary of the MAR is

$$K = (1 - \pi k_{Am} H)/(1 - \pi k_{Am} H), \qquad (9)$$

where $k_{Am} = \omega/V_{Am}$ is the wave number of the Alfven wave at the boundary of the MAR and $H$ is a characteristic scale of the decrease in this wave number when the wave propagates from the apex to the MAR boundary. The Eq. (9) is similar to that presented by Polyakov and Rapoport (1981) for the IAR. For this case, $\omega \approx$ 0.2-03 rad/s, $V_{Am} \approx$ 6.5 $10^9$ cm/s, $H \approx 1.25R_E$, and $K$ = 0.8-0.85. The reflected wave takes about 80-85% of energy of the incident wave and only 15-20% penetrates into the ionosphere. The reflected wave can propagate either through the barium cloud or beside it. The latter can be attributed, for example, to that the drift velocity of the plasma cloud with the diamagnetic cavity inside differs from the drift velocity of the Alfven wave in an external electric field. If the wave intersects the plasma cloud, it can be completely trapped by the cloud which is a specific resonator itself. This conclusion follows from the Eq. (9) for typical densities of the plasma cloud at $t <$ 100 s. We suppose that the wave initially emitted to the northern hemisphere and reflected once from the boundary of the MAR becomes trapped by the plasma cloud and leaves there all its energy heating the plasma of the cloud. On the contrary, the wave initially emitted to the southern hemisphere will never intersect the plasma cloud and loses its energy in the regions of reflection. This assumption is conditioned by that the distance $h_0 - h_S$ is approximately twice as the distance $h_0 - h_N$. We will further consider only this weakly decaying wave. The propagation of this wave from one reflection point to another takes about 35-40 s. The amplitude of this wave decreases by a factor of 4 upon 6-8 reflections that is approximately upon 4-6 min after the generation.

ACCELERATION OF ELECTRONS

Introduce an effective longitudinal current $J_{eff} = e\, n_e\, v_{Te}$. Near the generation region, the thermal velocity of electrons is $v_{Te}$ = 4 $10^7$ cm/s, $n_e$ = 2 cm$^{-3}$, and $J_{eff} \approx$ 1.3 $10^{-7}$ A/m$^2$. Since the electrons carry the longitudinal current of the Alfven wave, $J_z = e\, n_e\, v_d$ where $v_d$ is the drift velocity of electrons in the Alfven wave. When $J_z > J_{eff}$, the drift velocity of electrons exceeds their thermal velocity ($v_d > v_{Te}$) and the plasma instabilities occur. Strengthening the amplitude of plasma waves can give rise to a plasma turbulence and anomalous resistance. An anomalous collision frequency, in turn, gives rise to a longitudinal electric field $E_z$ (see, for example, Sagdeev and Galeev, 1969; Liperovsky and Pudovkin, 1983 ). Inside the magnetospheric Alfven resonator, $n_e = n(H^+)$. The condition $T_e/T_i > 3$ that is necessary for development of the ion-acoustic instability is not always fulfilled. Therefore, the electrostatic ion-cyclotron instability (EIC) is more probable. The growth rate of the EIC is $\gamma \sim \Omega(H^+)$, its threshold is $v_C \approx 0.4\, v_{Te}$, and the characteristic frequency of excitation is $\Omega \approx 1.5\, \Omega(H^+)$ (see, for example, Liperovsky and Pudovkin, 1983). The growth of the EIC wave amplitude gives rise to the EIC turbulence and anomalous resistance. For the developed EIC turbulence, the anomalous collision frequency is $v_a \approx 5\, \Omega(H^+)$ (Liperovsky and Pudovkin, 1983). For these conditions, we can estimate the longitudinal electric field using the equation (Sagdeev and Galeev, 1969)

$$E_z \approx (4\pi v_a / \omega_p^2)\, J_z, \qquad (10)$$

where $\omega_p$ is the electron plasma frequency and $J_z$ is the longitudinal current of the Alfven wave. The Eq. (10) shows that for a fixed $J_z$, the longitudinal field reaches its maximum near the boundaries of the MAR i.e. at the altitudes $h \approx R_E$. For this height, $B_0 = 6\ 10^{-2}$ G, $n_e = 4$ cm$^{-3}$, $\Omega(H^+) = 5.8\ 10^2$ rad/s and $E_z = 3.4$ mV/m, if we assume for estimates $J_z = 1\ 10^{-7}$ A/m$^2$ that somewhat exceeds the threshold. Let the $Z_T$ denote a characteristic longitudinal scale of the turbulent layer containing the field $E_z$. Apparently, the anomalous skin length of the Alfven waves (Lysak and Dum, 1983; Trakhtengertz and Feldshtein, 1987) determines this scale. For our case, $Z_T = 400\text{-}500$ km according to estimates. Coming into the turbulent layer, the suprathermal and energetic electrons will be accelerated along the magnetic field. They gain the energy

$$W_e = e\, E_z\, Z_T. \qquad (11)$$

crossing the turbulent layer. For the conditions considered, $W_e \approx 1.3\text{-}1.7$ keV. Since the Alfven wave contains the longitudinal currents of both directions, the electrons will be accelerated both toward the apex of the field line and toward the ionosphere. The electrons accelerated toward the ionosphere will change the ionospheric conductivity and give rise to a diffuse aurora.

Estimate the time intervals in which electrons can gain such energies assuming that the longitudinal threshold current is $J_z \approx 1\ 10^{-7}$ A/m$^2$. As mentioned above, the amplitude of the longitudinal current $J_m \approx J_R$. Figure 1b shows that the longitudinal current exceeds the threshold at $t > 20$ s. The maximum amplitude of the longitudinal current in the Alfven wave impulse exceeds the threshold approximately by a factor of 4. As indicated above, the amplitude of the Alfven wave takes a quarter of the initial amplitude only upon 6-8 oscillations between the boundaries of the MAR. It means that the acceleration of electrons could occur within the interval $0.5 < t < 4$-6 min after the release. This acceleration is virtually continuous for the first minute becoming then discrete with repetitions every 35-40 s due to processes at both boundaries of the MAR. It is difficult to estimate the time of aurorae related to the acceleration of electrons. Apparently, this time approximately corresponds to the 3-6th min after the release. It was mentioned above that electrons are also accelerated toward the apex of the field line. These electrons will oscillate between the mirror points without precipitation. After each acceleration session, the number of such electrons can increase. Eventually, these electrons can precipitate giving rise to a really observed aurora. Note that, according to the G-10 experiment, the first aurorae were observed upon 4 min after the release.

The acceleration of electrons at altitudes $h = R_E$ will be accompanied by the auroral kilometric radiation (AKR) at frequencies ~ 100kHz (see, for example, Lyons and Williams, 1984). The AKR can be observed as a continuous radiation during the first minute (starting upon 20 s after the release) and then in a discrete manner with pauses of 35-40 s till 4-6 min after the release. Apparently, this result does not contradict the G-10 experiment.

DISCUSSION

Our analysis of the processes associated with the Alfven wave is very approximate and qualitative. The most significant approximation was used in modeling the Alfven wave generation. Namely, the external source $J_R$ was considered as independent of the Alfven wave that is we neglected the back action of the Alfven wave on the source. Moreover, the formation of the diamagnetic cavity and plasma envelope was modeled in a very simplified form. In particular, we neglected the Hall term of the MHD equations which involves a development of strong irregularities in both the cavity and plasma envelope considerably complicating the phenomenon (Huba et al., 1992) Nevertheless, our findings do not contradict some results of the G-10 experiment. In particular, the relation of the AKR to generation and propagation of the Alfven wave is, apparently, the most natural.

We cannot explain the processes observed upon 5-6 min after the release using only the Alfven wave model. At the same time, these processes were rather intense and resulted in aurorae clearly observed approximately upon 8-10 min after the release (Bernhardt, 1992) and ultra short radio wave storm at frequencies 50-300 MHz (Oraevsky et al., 1995). The electrons of relatively high energies 1-10 keV are necessary for these phenomena. It is plausible that such electrons can be attributed to the well known betatron heating (see, for example, Akasofu and Chapman, 1972) which occurs at the collapse of the diamagnetic cavity that is a recovery of the magnetic field to its background value $B_0$. The spherical diamagnetic cavity created by the expansion of the plasma cloud reached its maximum radius $R_B \approx 35$ km at $t = t_B \approx 0.6$ min. Denote the minimum magnetic field in the cavity for $t = t_B$ as $B_B$. The G-10 experiment indicates that $B_0/B_B > 10^2$ (Bernhardt, 1992; Huba et al.,1992). For our estimates, assume that $B_0/B_B = 10^3$.

Since the diamagnetic cavity could be topologically disconnected from the lines of force of the geomagnetic field, we suppose that the inner plasma cannot escape from the cavity. For this case, the lifetime of the diamagnetic cavity is determined by the diffusion of magnetic field

$$\tau_m = 4 \pi \sigma R_B^2/c^2, \tag{12}$$

where $\sigma$ is the plasma conductivity inside the cavity. In turn, the plasma conductivity is determined by the Coulomb collisions. Therefore, the conductivity is independent of the plasma concentration but increases with growth in the electron temperature $T_e$: $\sigma = e^2 n_e/(m_e \nu_e)$ ; $\nu_e \sim n_e/T_e^{3/2}$. In our case, $\tau_m \approx 8\text{-}10$ min since we suppose that the aurorae are attributed to the collapse of diamagnetic cavity. This permits to estimate the electron temperature $T_{eB}$ for $t = t_B$ using the formula (12). Thus we obtain $T_{eB} \sim 10^{-2}$ eV. Such temperature is quite plausible, if we take into account cooling the inner plasma under the expansion (note that the real temperature can be higher since the lifetime is also determined by the processes of plasma escape). In addition to thermal (cold) electrons with the temperature $T_{eB}$, there can be suprathermal electrons in the cavity, for example the photoelectrons with energies 5-10 eV originating due to the continuous photoionization of neutrals. For these relatively hot electrons, the characteristic time between the Coulomb collisions is about 500 s. Therefore, these initially suprathermal electrons can be accelerated to the energies 5-10 keV via the betatron mechanism. The possible initial concentration of such electrons can be about 10-100 cm$^{-3}$ since the number density of cold electrons exceeds $10^3$ cm$^{-3}$. The concentration and energy of these electrons can be quite sufficient for the observed aurorae. Note that the electrons of the same energies were invoked in explanation of the ultra short radio wave storm (Oraevsky et al., 1995).

CONCLUSIONS

The qualitative study of the processes resulting from the barium release at altitudes $\sim 5.2\ R_E$ shows that the generation of Alfven waves is possible during the total time of the expansion of plasma cloud. The maximum intensity of these waves corresponds to the time of the complete retardation of the diamagnetic cavity created by the expansion of the plasma cloud. For conditions relevant to the G-8 and G-10 experiments of the CRRES mission, the Alfven wave exhibits the form of an impulse with an effective frequency $\sim 0.03\text{-}0.05$ Hz. Due to the background conditions and wave frequency, the wave mainly oscillates along the geomagnetic field between the mirror reflection points situated at $\sim 0.7\ R_E$ i.e. the wave is trapped by the magnetospheric Alfven resonator. The reflection coefficient is about 80-85%. The wave amplitude is sufficient to the generation of plasma turbulence, the longitudinal electric field, and to an increase in the longitudinal energy of electrons accelerated by this field up to $\sim 1$ keV. These processes are the most probable for altitudes $\sim 1\ R_E$. The auroral kilometric radiation (AKR) at frequencies $\sim 100$ kHz is associated with these accelerated electrons. It is shown that the acceleration of electrons and AKR can be observed almost continuously during the first minute and then from time to time with pauses about 35-40 s till 6-8 min after the release. Apparently, these findings do not contradict the experimental data. The betatron acceleration of electrons at the recovery of the geomagnetic field is also discussed. This mechanism could be responsible for the acceleration of electrons resulting in the aurorae and ultra short radio wave storm at frequencies 50-300 MHz observed at the 8-10th min after the release.

ACKNOWLEDGMENT

V.I. Badin acknowledges the support of the Russian Foundation for Basic Research, project no. 99-05-65080.

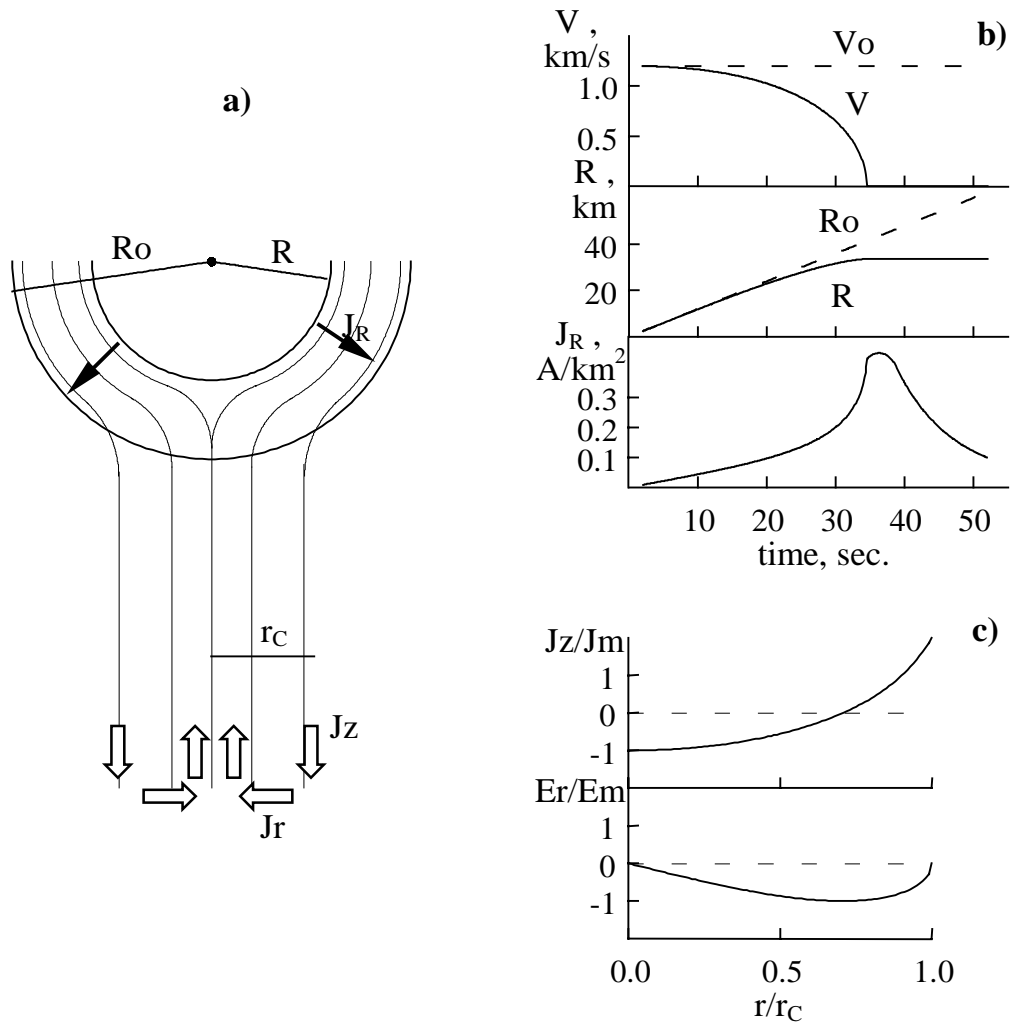

Fig. 1a. The qualitative scheme of the radial current $J_R$ inside the plasma envelope (between the diamagnetic cavity of the radius $R$ and the boundary of the plasma cloud with the radius $R_0$), and the longitudinal $J_z$ and radial $J_r$ currents of the Alfven wave. Thin curves are the magnetic field lines, $r_C$ is the radial width occupied by the Alfven wave.

Fig. 1b. The characteristic features of the plasma envelope depending on time following the barium release: $V$ is the radial velocity of the expansion of the diamagnetic cavity and magnetic field lines surrounding this cavity, $V_0$ is the radial velocity of barium ions immediately after ionization of barium atoms.

Fig 1c. The qualitative scheme of the radial distribution of the longitudinal current $J_z$ and transversal electric field $E_r$ of the Alfven wave at a fixed time. The $J_m$ and $E_m$ are their maximum amplitudes.